\documentclass[
    ,final            % use final for the camera ready runs
  ]
  {aipproc}

\layoutstyle{6x9}

\usepackage{xspace}
\usepackage[scaled=0.92]{helvet}
\usepackage{upgreek}
\usepackage{psfrag}

\newcommand{\hb}{\mbox{HERA--B}\xspace}
\def\babar{\mbox{\slshape B\kern-0.1em{\smaller A}\kern-0.1em B\kern-0.1em{\smaller A\kern-0.1em R}}}

\newcommand{\pA}{\ensuremath{{\mathrm{p}A}}}
\newcommand{\pN}{\ensuremath{{\mathrm{p}N}}}

\newcommand{\eps}{\ensuremath{\varepsilon}\xspace}

\newcommand{\eg}{e.g.\ }

\newcommand{\ra}{\ensuremath{\rightarrow}}

\newcommand{\unit}[1]{\ensuremath{\:\mathrm{#1}}}
\newcommand{\statsyst}[2]{\ensuremath{\pm#1\mathrm{(stat.)}\pm#2\mathrm{(syst.)}}}

\newcommand{\gev}{\unit{GeV}}

\newcommand{\xf} {{\ensuremath{x_\mathrm{F}}}\xspace}
\newcommand{\pt} {{\ensuremath{p_\mathrm{T}}}\xspace}

\newcommand{\cquark}{{\ensuremath{\mathrm{c}}}\xspace}
\newcommand{\bquark}{{\ensuremath{\mathrm{b}}}\xspace}

\newcommand{\PJgy}{{\ensuremath{\mathrm{J}\mskip -2mu/\mskip -2mu\uppsi}}\xspace}
\newcommand{\jpsi}{\PJgy}

\newcommand{\psitwos}{{\ensuremath{\uppsi(2S)}}\xspace}
\newcommand{\chic}{{\ensuremath{\chi_\cquark}}\xspace}

\newcommand{\bbbar}{{\ensuremath{\mathrm{b\overline{b}}}}\xspace}

\newcommand{\dilepton}{{\ensuremath{\ell^+ \ell^-}}}               % lepton pair

\newcommand{\epem}{{\ensuremath{\mathrm{e^+ e^-}}}}                         % e+ e-
\newcommand{\dielectron}{\epem\xspace}
\newcommand{\mpmm}{\ensuremath{\upmu^+ \upmu^-}\xspace}                  % mu+ mu-
\newcommand{\dimuon}{\mpmm}
\newcommand{\jpmm}{{\ensuremath{\jpsi\rightarrow\dimuon}}\xspace}
\newcommand{\jpsimumu}{\jpmm}
\newcommand{\jpee}{{\ensuremath{\jpsi\rightarrow\dielectron}}\xspace}
\newcommand{\jpsiee}{\jpee}

\DeclareFontFamily{OML}{mygreek}{\skewchar \font =127}
\DeclareFontShape{OML}{mygreek}{m}{rm}{<-> [0.91] zptmcm7m }{}
\DeclareSymbolFont{lettersgreek}{OML}{mygreek}{m}{rm}

\DeclareMathSymbol{\alpha}{0}{lettersgreek}{"0B}
\DeclareMathSymbol{\beta}{0}{lettersgreek}{"0C}
\DeclareMathSymbol{\gamma}{0}{lettersgreek}{"0D}
\DeclareMathSymbol{\delta}{0}{lettersgreek}{"0E}
\DeclareMathSymbol{\epsilon}{0}{lettersgreek}{"0F}
\DeclareMathSymbol{\zeta}{0}{lettersgreek}{"10}
\DeclareMathSymbol{\eta}{0}{lettersgreek}{"11}
\DeclareMathSymbol{\theta}{0}{lettersgreek}{"12}
\DeclareMathSymbol{\iota}{0}{lettersgreek}{"13}
\DeclareMathSymbol{\kappa}{0}{lettersgreek}{"14}
\DeclareMathSymbol{\lambda}{0}{lettersgreek}{"15}
\DeclareMathSymbol{\mu}{0}{lettersgreek}{"16}
\DeclareMathSymbol{\nu}{0}{lettersgreek}{"17}
\DeclareMathSymbol{\xi}{0}{lettersgreek}{"18}
\DeclareMathSymbol{\pi}{0}{lettersgreek}{"19}
\DeclareMathSymbol{\rho}{0}{lettersgreek}{"1A}
\DeclareMathSymbol{\sigma}{0}{lettersgreek}{"1B}
\DeclareMathSymbol{\tau}{0}{lettersgreek}{"1C}
\DeclareMathSymbol{\upsilon}{0}{lettersgreek}{"1D}
\DeclareMathSymbol{\phi}{0}{lettersgreek}{"1E}
\DeclareMathSymbol{\chi}{0}{lettersgreek}{"1F}
\DeclareMathSymbol{\psi}{0}{lettersgreek}{"20}
\DeclareMathSymbol{\omega}{0}{lettersgreek}{"21}
\DeclareMathSymbol{\varepsilon}{0}{lettersgreek}{"22}
\DeclareMathSymbol{\vartheta}{0}{lettersgreek}{"23}
\DeclareMathSymbol{\varomega}{0}{lettersgreek}{"24}
\DeclareMathSymbol{\varrho}{0}{lettersgreek}{"25}
\DeclareMathSymbol{\varsigma}{0}{lettersgreek}{"26}
\DeclareMathSymbol{\varphi}{0}{lettersgreek}{"27}

\begin{document}

\title{Charm and Beauty Production at \hb}

\classification{13.20.Gd, 13.20.He, 13.85.Qk,14.40.Gx, 24.85.+p}
\keywords      {Heavy quarks, \jpsi production, nuclear effects, beauty production}

\author{Ulrich Husemann for the \hb Collaboration}{%
address={Fachbereich Physik, Universit\"at Siegen, D--57068 Siegen, Germany\\
now at University of Rochester, Rochester, New York 14627}%
}

%\author{<author2>}{
%  address={<common address for author2 and author3>}
%}

%\author{<author3>}{
%  address={<common address for author2 and author3>}
%  ,altaddress={<author1 address>} % additional visiting address
%}

\begin{abstract}
The \hb experiment at DESY has acquired a data-set of approximately
300,000 decays $\jpsi\rightarrow\dilepton$ during
its 2002/2003 data-taking period. These data are used to analyze
the production of heavy quarks in proton-nucleus interactions at
a center-of-mass energy of 41.6\gev.
In this article, preliminary results of two measurements are discussed,
a measurement of nuclear effects in the production of \jpsi mesons and
a measurement of the \bbbar production cross section.

\end{abstract}

\maketitle

%%%%%%%%%%%%%%%%%%%%%%%%%%%%%%%%%%%%%%%%%%%%
%% MAINMATTER
%%%%%%%%%%%%%%%%%%%%%%%%%%%%%%%%%%%%%%%%%%%%

%\section{Introduction}

\section{\hb: Detector, Trigger and Data-Set}

\begin{figure}[t]
\includegraphics[width=0.60\textwidth]{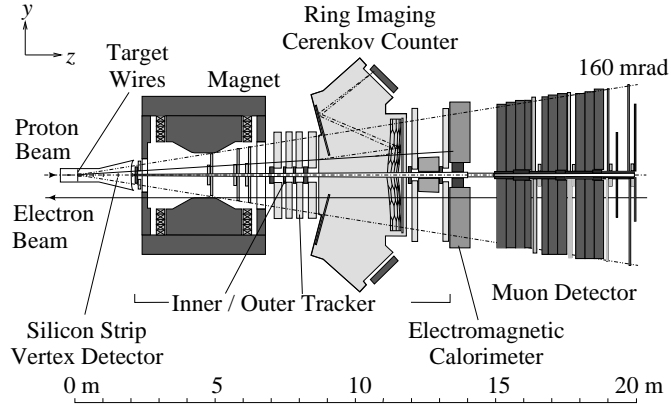}
\caption{Elevation view of the \hb detector.}
\label{fig:herab}
\end{figure}

The \hb detector, depicted in Fig.~\ref{fig:herab}, is a fixed-target 
spectrometer with large angular acceptance. Protons from the
halo of the HERA proton beam are collided with an internal wire
target. The target consists of up to eight thin wires of different materials,
distributed in two target stations. Each wire can be moved independently 
to adjust the interaction rate.

The \hb detector consists of the following sub-detectors: 
A silicon micro-strip vertex detector is used to reconstruct and separate 
primary and secondary vertices. The tracking detectors comprise 
micro-strip gaseous chambers with gas electron multiplier foils 
in the inner acceptance region of high particle flux
and honeycomb drift chambers in the outer region.
Particle identification is performed by a ring-imaging
\v Cerenkov counter, an electromagnetic calorimeter, and a four-layer
muon detector.

Lepton pairs from \jpsi decays are enriched by a multi-level trigger system.
The first trigger level is implemented as a hardware track trigger in the outer
tracking detector. Starting points for the track search are provided 
by pretriggers in the muon detector and the calorimeter. At the second
trigger level, a software trigger running on a PC farm, track candidates 
are extrapolated to the vertex detector, 
and a two-prong vertex fit is performed. Accepted events are reconstructed
online.

The \hb experiment has taken data between October 2002 and February 2003.
The data sample recorded with the lepton pair trigger 
contains approximately 150,000 decays into each of the 
channels \jpsiee and \jpsimumu, 90,000 of which were taken with
two target wires of different materials simultaneously.
Based on this data set, the \hb collaboration has analyzed the production
of charmonium states such as \jpsi, \psitwos and \chic, as well as the
production of open and hidden beauty, \eg \bbbar and $\Upsilon$.

%\vspace{-2mm}

\section{Nuclear Effects in \jpsi Production}
%\vspace{-2mm}

Many theoretical models of \jpsi production in proton-nucleus interactions
predict modifications of the production cross section by nuclear effects,
\eg due to the absorption of final state particles in the nuclear
environment~\cite{Vogt:1999dw,Vogt:2001ky}, or due to coherent proton-nucleus 
scattering~\cite{Kopeliovich:2001ee,Boreskov:2003ck}.
The nuclear effects are commonly parametrized by 
the power law
\begin{equation}
  \sigma_{\pA} = \sigma_{\pN} \cdot A^{\alpha(\xf,\pt)}.
  \label{eq:suppression}
\end{equation}
Here, $\sigma_{\pA}$ is the proton-nucleus cross section, $\sigma_{\pN}$ is the
proton-nucleon cross section, and $A$ is the atomic mass number of the target.
A value of the nuclear suppression parameter 
\mbox{$\alpha(\xf,\pt)$} smaller than unity corresponds to 
the suppression of \jpsi production. 
The cross section $\sigma$ is related to the number of
detected particles $N$, the integrated luminosity $\mathcal{L}$, and
the detection efficiency \eps by $\sigma=N/(\mathcal{L}\varepsilon)$.
Using this relation for
a tungsten target and a carbon target operated at the same time,
Eq.~(\ref{eq:suppression}) can be solved for~$\alpha$:
\begin{equation}
  \alpha = \frac{1}{\log\left(A_\mathrm{W}/A_\mathrm{C}\right)}
  \cdot \log\left( 
    \frac{N_\mathrm{W}}{N_\mathrm{C}}\cdot
    \frac{\mathcal{L}_\mathrm{C}}{\mathcal{L}_\mathrm{W}}\cdot
    \frac{\varepsilon_\mathrm{C}}{\varepsilon_\mathrm{W}}
  \right).
\end{equation}
The measurement hence includes measuring three ratios, the
ratio of \jpsi yields, the ratio of luminosities per wire,
and the ratio of efficiencies. The analysis thus
benefits from the advantages of a relative measurement, in which many
systematic uncertainties cancel out. The \jpsi yields
are determined by fits to the \jpsi invariant mass spectrum in bins of the
kinematic variables. To measure the luminosities, the number of primary 
vertices per target wire is counted in data recorded with
a zero-bias trigger in parallel to the lepton pair trigger.
A detailed Monte Carlo (MC) simulation of the \hb detector and trigger is
performed to calculate the ratio of efficiencies.
Details of the analysis in the decay channel \jpsimumu 
can be found in~\cite{Husemann:2005}.
\enlargethispage{2mm}

\begin{figure}[t]
%\subfigure[]{
\includegraphics[width=0.38\textwidth]{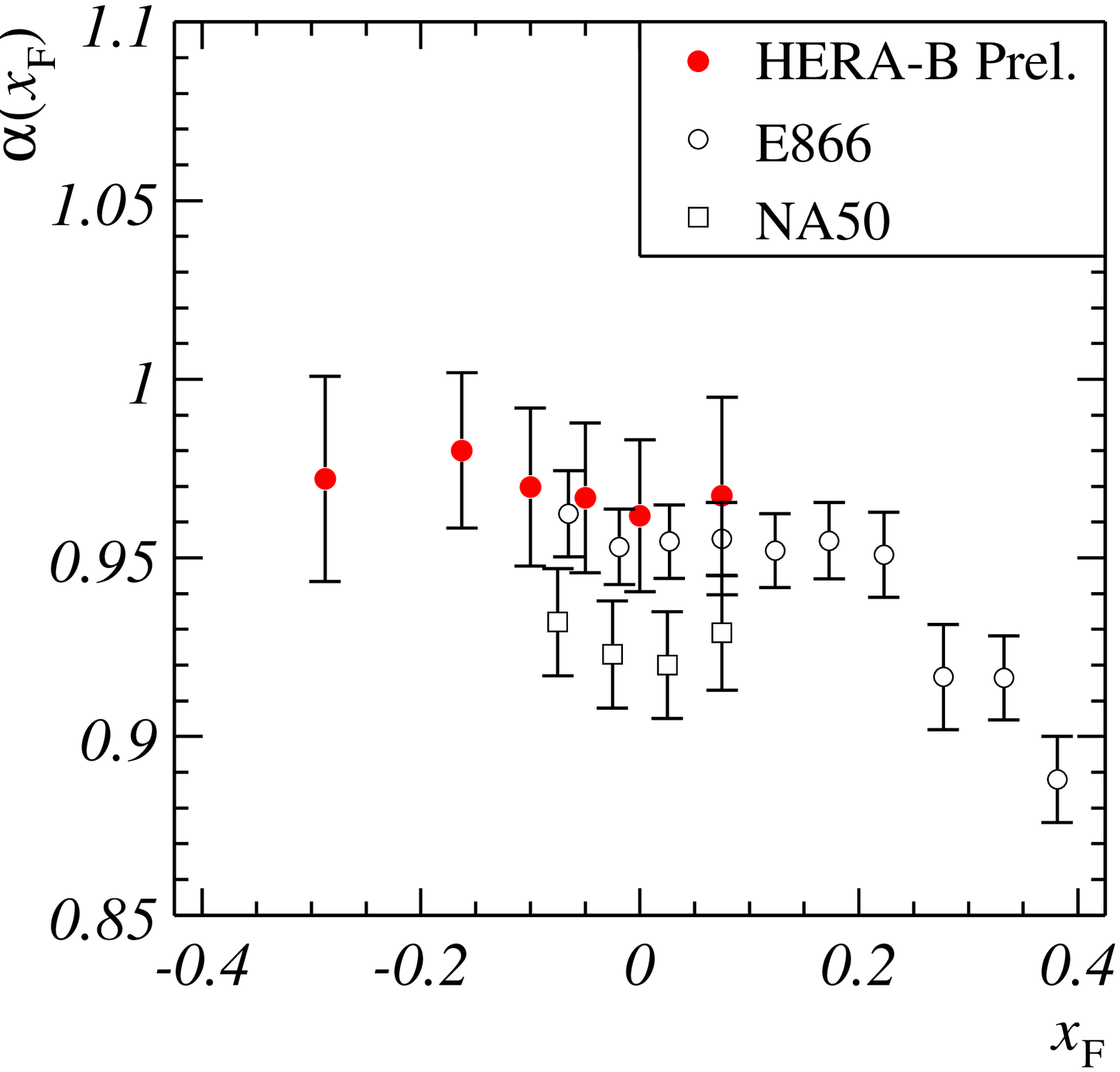}%}
\hspace{0.1\textwidth}
%\subfigure[]{
\includegraphics[width=0.38\textwidth]{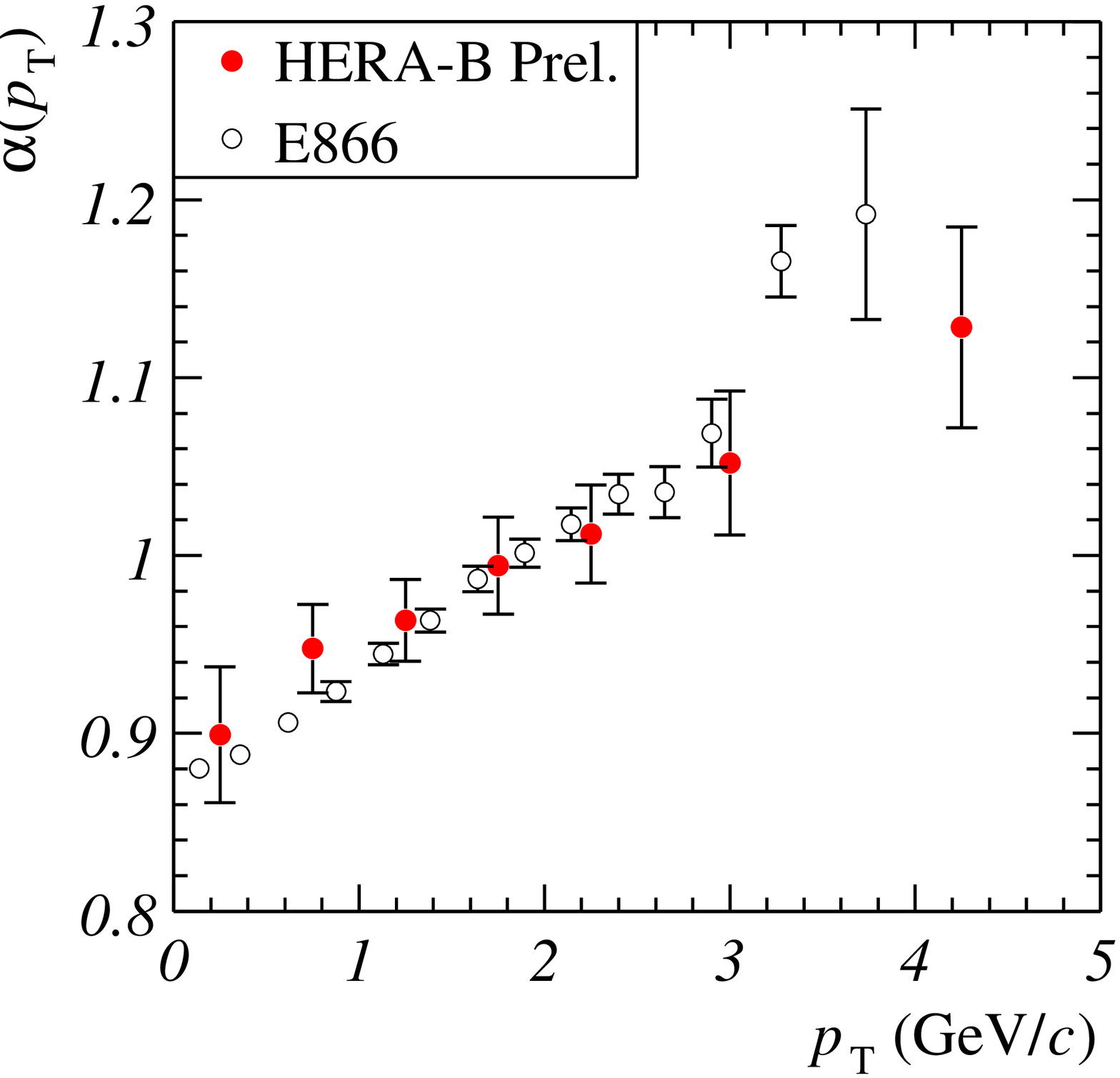}
%}
\caption{Nuclear suppression parameter $\alpha$ in the decay channel 
\jpsimumu as a function of \xf~(left) and \pt~(right), compared to previous
measurements by the E866~\cite{Leitch:1999ea} and the 
NA50~\cite{Alessandro:2003pc} collaborations.
All data points include
statistical and systematic uncertainties.}
\label{fig:adep}
\end{figure}

The measured nuclear suppression parameter $\alpha$ in the decay channel
\jpsimumu is shown in Fig.~\ref{fig:adep} as a function of \xf and \pt. 
The data show good agreement
with previous measurements in the overlap region and extend the existing data
to negative \xf. In the covered range of $-0.375\leq \xf < 0.125$, 
a small constant suppression of \jpsi production is observed, with an 
average suppression parameter of 
%\begin{equation}
$
\overline{\alpha} = 0.969\statsyst{0.003}{0.021}.
$
%\end{equation}

\section{The $\mathbf{b\overline{b}}$ Production Cross Section}
At the \hb center-of-mass energy of 41.6\gev, beauty production is close
to its kinematic threshold. This results in large theoretical uncertainties
due to the re-summation of soft gluons and increased 
sensitivity to the mass of the
\bquark quark. Previous measurements of the \bbbar production cross section 
in a similar energy range suffered from 
small statistics and large systematic uncertainties.

In \hb, the \bbbar production cross section is measured in the inclusive
decay channel $\bbbar\ra\jpsi\,X\ra\dilepton\,X$, where
$\ell=\mathrm{e},\upmu$. The average flight
distance of B mesons in the \hb detector amounts to approximately 8\unit{mm}.
Therefore, B candidate events can be selected based on the separation of the
\jpsi decay vertex from the primary vertex on the target wire. The best
separation from background decays such as prompt \jpsi decays and semileptonic
\bquark and \cquark decays is achieved by a combined cut on the significance
of the \jpsi detached vertex and the impact parameter to the target wire.
Furthermore, combinatorial background is controlled by comparing the background
upstream and downstream of the target with respect to the proton flight 
direction. A~fit to the B lifetime and the requirement of a third track at
the \jpsi decay vertex serve as independent confirmations of the B flavor.
For the final \bbbar cross section result, the consistent results in the
dielectron and 
the dimuon channels using the 2002/2003 data-sample are combined with the 
previously published result~\cite{Abt:2002rd} from the 2000 data-taking.

\begin{figure}[t]
  \psfrag{s}[r][]{\small $\sigma_{\bbbar}$\,(nb/nucl.)}
  \psfrag{p}[r][]{\small proton energy (GeV)}
  \psfrag{prel}[r][]{\small \hb Preliminary}
  \includegraphics[width=0.45\textwidth]{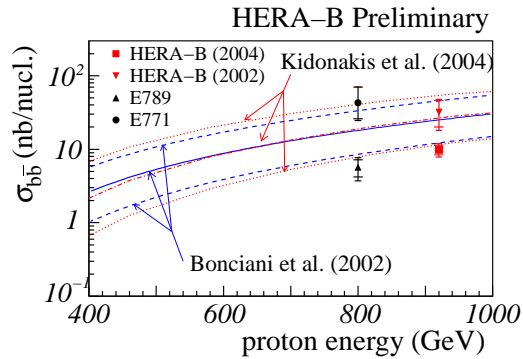}
  \caption{Preliminary cross section for \bbbar production as a function of
    the proton energy, compared to theoretical 
    predictions~\cite{Bonciani:1998vc,Kidonakis:2004qe} and a previous
    \hb measurement based on the 2000 data-set~\cite{Abt:2002rd}.}
  \label{fig:bb_theory}
\end{figure}

To minimize the influence of external input, the result of the measurement 
is first presented relative to the \jpsi production cross section and 
restricted to the kinematic acceptance of the \hb detector:
\begin{equation}
R_\bbbar = \frac{\Delta\sigma_\bbbar}{\Delta\sigma_\jpsi}
= \eps\, \frac{N_{\bbbar\ra\jpsi\,X}}{N_\mathrm{prompt\,\,\jpsi}}
= 0.033\statsyst{0.005}{0.004}.
\end{equation}
Here, $\Delta\sigma_\bbbar$ and $\Delta\sigma_\jpsi$ are the production
cross sections for \bbbar and \jpsi within the \hb acceptance,
$N_{\bbbar\ra\jpsi\,X}$ and $N_\mathrm{prompt\,\,\jpsi}$ are the number of
\bbbar candidates and the number of prompt \jpsi. The efficiency parameter
\eps includes the branching fractions, the efficiency of the detached vertex
selection and the nuclear dependences for \bbbar and \jpsi events. 
The efficiencies are determined from a detailed MC simulation. 
The result is extrapolated to the full phase-space with the help
of PYTHIA~\cite{Sjostrand:1994yb}. Since to date, 
there is no published \hb measurement of the \jpsi production cross 
section, a value of $\sigma_\jpsi = 352\statsyst{2}{26}\unit{nb/nucl.}$ 
is assumed~\cite{Abt:2002rd}. 
This value is obtained by averaging $\sigma_\jpsi$ as 
measured by the
E789 and E771 experiments~\cite{Schub:1995pu,Alexopoulos:1995dt} 
and extrapolating the result to \hb energies.
Note that a significant upward revision of this value is anticipated 
from the \hb measurement.
The resulting \bbbar production cross section is depicted in
Fig.~\ref{fig:bb_theory}, along with previous measurements and theoretical
predictions. The preliminary numerical result is
%\begin{equation}
$
\sigma_\bbbar = 9.9\statsyst{1.5}{1.4} \unit{nb/nucl.}
$
%\end{equation}

%\section{Summary and Outlook}
%In this article, measurements of the nuclear dependence of \jpsi production 
%and of \bbbar production are presented.
%Apart from these analyses, \hb has achieved further
%preliminary results on heavy flavor production,
%including measurements of the \jpsi production cross section, \chic and
%\psitwos production, and a measurement of the $\Upsilon$ production cross
%section.

%%%%%%%%%%%%%%%%%%%%%%%%%%%%%%%%%%%%%%%%%%%%%%%%
%% BACKMATTER
%%%%%%%%%%%%%%%%%%%%%%%%%%%%%%%%%%%%%%%%%%%%%%%%
%\vspace{-2mm}
\enlargethispage{5mm}

\begin{theacknowledgments}
The author would like to thank the DIS05 crew
for the well-organized and interesting conference. 
This work was supported by the German Bundesministerium f\"ur 
Bildung und Forschung under the contract number 5HB1PEA/7.
Special thanks to the
\hb{} charmonium and beauty working groups for their help in preparing 
this conference contribution and to DESY for the kind financial support.
\end{theacknowledgments}

%%%%%%%%%%%%%%%%%%%%%%%%%%%%%%%%%%%%%%%%%%%%%%%%
%% The bibliography can be prepared using the BibTeX program or
%% manually.
%%
%% The code below assumes that BibTeX is used.  If the bibliography is
%% produced without BibTeX comment out the following lines and see the
%% aipguide.pdf for further information.
%%
%% For your convenience a manually coded example is appended
%% after the \end{document}
%%%%%%%%%%%%%%%%%%%%%%%%%%%%%%%%%%%%%%%%%%%%%%%%

%%%%%%%%%%%%%%%%%%%%%%%%%%%%%%%%%%%%%%%%%%%%%%%%
%% You may have to change the BibTeX style below, depending on your
%% setup or preferences.
%%
%%
%% For The AIP proceedings layouts use either
%%%%%%%%%%%%%%%%%%%%%%%%%%%%%%%%%%%%%%%%%%%%

%\vspace{-2mm}

%\bibliographystyle{aipproc}   % if natbib is available
%\bibliographystyle{aipprocl} % if natbib is missing

%%%%%%%%%%%%%%%%%%%%%%%%%%%%%%%%%%%%%%%%%%%
%% You probably want to use your own bibtex database here
%%%%%%%%%%%%%%%%%%%%%%%%%%%%%%%%%%%%%%%%%%%
%\bibliography{dis}

%%%%%%%%%%%%%%%%%%%%%%%%%%%%%%%%%%%%%%%%%%%
%% Just a reminder that you may have to run bibtex
%% All of it up to \end{document} can be removed
%% if you don't like the warning.
%%%%%%%%%%%%%%%%%%%%%%%%%%%%%%%%%%%%%%%%%%%
\IfFileExists{\jobname.bbl}{}
 {\typeout{}
  \typeout{******************************************}
  \typeout{** Please run "bibtex \jobname" to optain}
  \typeout{** the bibliography and then re-run LaTeX}
  \typeout{** twice to fix the references!}
  \typeout{******************************************}
  \typeout{}
 }

\end{document}